\documentclass[aps, pre, twocolumn, 10pt]{revtex4-2}

\usepackage{amsfonts,amsmath,bm,amssymb,color}
\usepackage{graphicx}
\usepackage[colorlinks=true, citecolor=blue, linkcolor=blue]{hyperref}
\usepackage{subcaption}
\usepackage[table]{xcolor}
\usepackage{colortbl}

\begin{document}
\title{Bounds and sufficient conditions in quantum Zeno effect}

\author{Vladimir Parfenyev$^{1,2}$ and Sergey Belan$^{1,2}$}\email{sergb27@yandex.ru} 
\affiliation{$^1$Landau\ Institute\ for\ Theoretical\ Physics,\ Russian\ Academy\ of\ Sciences,\ 1-A Akademika\  Semenova av.,\ 142432\ Chernogolovka,\ Russia}
\affiliation{$^2$HSE\ University,\ Faculty\ of\ Physics,\ Myasnitskaya\ 20,\ 101000\ Moscow,\ Russia}

\begin{abstract}
When a quantum system undergoes a series of repeated measurements during its time evolution to check whether it remains in the initial state, the resulting state lifetime can decrease (the quantum Zeno effect) or increase (the quantum inverse Zeno effect) as the inter-measurement period increases. Several transitions between these two regimes may occur within the same system. A carefully chosen measurement period $\tau_\ast$ (or rate $r_\ast$ in the case of a stochastic measurement protocol) brings the state lifetime to its global minimum. Here, we describe the upper and lower bounds on the optimized state lifetime, which are expressed in terms of characteristic time scales encoded in the state survival probability $P(T)$, which is assumed to decrease monotonically to zero at large $T$. In addition, we analyze the bounds on the optimal measurement periods/rates and formulate sufficient conditions for observing the inverse Zeno effect. The analysis is also extended to states with non-monotonic survival functions. The results are illustrated numerically with an example of a system of optically trapped ultra-cold atoms.
\end{abstract}

\date\today

\maketitle

\section{Introduction}\label{sec: introduction}

Repeated measurements of an unstable quantum system can inhibit its time evolution, giving rise to the quantum Zeno effect. First recognized in the early days of quantum theory, this phenomenon has since been investigated in a wide range of experimental settings~\cite{itano1990quantum,kwiat1995interaction,streed2006continuous,helmer2009quantum,wolters2013quantum,patil2015measurement,kakuyanagi2015observation,slichter2016quantum,kalb2016experimental}. In the idealized limit of infinitely frequent, instantaneous measurements, the decay can be completely suppressed, effectively freezing the system in its initial state. Conversely, repeated measurements can also accelerate quantum decay, an effect known as the inverse Zeno effect, which has likewise been theoretically investigated and experimentally explored~\cite{kofman2000acceleration,facchi2001quantum,fischer2001observation,harrington2017quantum}. The coexistence of the Zeno and inverse Zeno regimes within a single quantum system suggests the possibility of controlling quantum transition rates through an appropriately designed measurement protocol. More broadly, quantum Zeno dynamics forms part of the wider research on measurement-based control of quantum systems~\cite{Smirnov_2017,Briegel_2009,Gundin_2025,Liu_2010,Leppenen_2021,Bednorz_2012,Leppenen_2022,Zibinskiy_2026,Nedelea_2023}.

From a theoretical perspective, quantum Zeno dynamics can be studied using two distinct approaches. The first focuses on how specific measurement protocols affect the state lifetime in particular quantum systems; see, e.g., Ref.~\cite{chaudhry2014zeno}. The second aims to establish general results concerning the influence of arbitrary measurement protocols on the lifetimes of arbitrary quantum states~\cite{belan2020optimality}. Within the latter framework, the results are formulated in terms of statistical properties of the state survival probability $P(T)$, rather than parameters specific to a particular physical system. Consequently, the resulting conclusions are robust against changes in the details of the underlying model. Both approaches provide valuable insights into how repeated measurements affect quantum transition rates. In particular, theoretical studies have established criteria for distinguishing between the quantum Zeno and inverse Zeno effects. For example, Ref.~\cite{zhang2018criterion} relates the occurrence of Zeno or inverse Zeno dynamics in a two-level system coupled to a dissipative environment to the convexity properties of the spectrum. Complementarily, the approach developed in Ref.~\cite{belan2020optimality} yields criteria for identifying the Zeno and inverse Zeno regimes in terms of the decay statistics at a given measurement frequency, without requiring detailed knowledge of the interaction Hamiltonian.

Here, we adopt the latter approach to establish some general properties of quantum Zeno and inverse Zeno dynamics under different measurement protocols. The survival probability $P(T)$ of an unstable quantum system is defined as the probability of finding the system in its initial state after an elapsed time $T$, provided that no measurements have been performed during this time. Suppose that the system is subjected to a sequence of repeated measurements at regular intervals $\tau$, a protocol known as stroboscopic, to determine whether it remains in its initial state. The characteristic decay time can then be naturally quantified by the average time required to detect the system's decay,
\begin{equation}
\label{mean_detection_time_str}
\langle t_\tau\rangle=\frac{\tau}{1-P(\tau)},
\end{equation}
as introduced in Ref.~\cite{belan2020optimality}. This expression assumes that the survival probability after multiple measurements factorizes, with each measurement modeled as an instantaneous projection. In other words, whenever a measurement fails to detect decay, the system's subsequent evolution begins anew from the same initial (undecayed) state.

The Zeno and inverse Zeno regimes can be distinguished by examining the dependence of $\langle t_\tau\rangle$ on the measurement period $\tau$. Specifically, the quantum Zeno effect occurs when $\partial_\tau \langle t_\tau\rangle < 0$, meaning that the mean decay detection time increases as the measurement period decreases. Conversely, when $\partial_\tau \langle t_\tau\rangle > 0$, the system exhibits inverse Zeno dynamics, with more frequent measurements leading to a shorter mean decay detection time. In the general case, the short-time behavior of the survival probability satisfies $1-P(\tau)\propto \tau^2$ as $\tau\to0$, see Ref.~\cite{ghirardi1979small}. Consequently, $\langle t_\tau\rangle\propto 1/\tau$ in this limit. At long times, $P(\tau)\to0$ as $\tau\to\infty$, and hence $\langle t_\tau\rangle\propto\tau$. These limiting behaviors imply that the mean decay detection time $\langle t_\tau \rangle$ has at least one minimum at a finite measurement period $\tau=\tau_\ast$, satisfying $\partial_\tau \langle t_\tau\rangle|_{\tau_\ast}=0$. In general, multiple transitions between the Zeno and inverse Zeno regimes may occur, corresponding to local extrema of $\langle t_\tau\rangle$ as a function of $\tau$. In what follows, $\tau_\ast$ will denote the location of the global minimum of $\langle t_\tau\rangle$.

Beyond strictly periodic stroboscopic measurements, several other measurement protocols have been investigated in the literature. In particular, when measurements are performed at random times~\cite{shushin2011effect,power1996stochastic,belavkin2000quantum}, with successive intervals drawn independently from a probability density $\rho(T)$, the mean detection time is given by
\begin{equation}
\label{stochastic_mean_detect_time}
\langle t\rangle=\frac{\langle T \rangle}{1-\langle P(T) \rangle}=\frac{\int_0^\infty dT \rho(T)T}{1-\int_0^\infty dT \rho(T)P(T)}.
\end{equation}
For the special choice $\rho(T)=\delta(T-\tau)$, corresponding to a fixed interval $\tau$ between successive measurements, this expression reduces to Eq.~\eqref{mean_detection_time_str}. 

When the intervals between successive measurements are exponentially distributed, with probability density $\rho(T)=re^{-rT}$, the measurements follow a Poisson process with rate $r$. Substituting this distribution into Eq.~\eqref{stochastic_mean_detect_time} yields \cite{belan2020optimality}
\begin{equation}
\label{poisson_mdt}
\langle t_{r}\rangle=\frac{1}{r(1-r\tilde P(r))},
\end{equation}
where $\tilde P(r)=\int_0^\infty dTe^{-rT}P(T)$ is the Laplace transform of the survival probability $P(T)$ evaluated at $r$. Depending on the sign of the derivative $\partial_r\langle t_r\rangle$ the system is either in Zeno ($\partial_r\langle t_r\rangle>0$) or inverse Zeno ($\partial_r\langle t_r\rangle<0$) regime with respect to Poisson measurements with rate $r$. 

Among all possible measurement protocols, stroboscopic measurements with an appropriately chosen period $\tau_\ast$ are optimal in the sense that they minimize the mean decay detection time, achieving its minimum at the transition between the quantum Zeno and inverse Zeno regimes~\cite{belan2020optimality}. More formally, if $\langle t_{\tau_\ast}\rangle\le \langle t_\tau\rangle$ for any $\tau>0$, then $\langle t_{\tau_\ast}\rangle\le \langle t_{\cal M}\rangle$ for any measurement protocol $\mathcal M$, whether deterministic or stochastic. This result holds universally, irrespective of the specific form of the survival probability $P(T)$.

Below, we derive several other general properties of Zeno and inverse Zeno dynamics in quantum decay problems that hold irrespective of the fine details of the particular class of quantum systems, assuming only that the survival probability $P(T)$ is a monotonically decreasing function that tends to zero as $T\to\infty$ (the case of non-monotonic $P(T)$ is briefly addressed in Section~\ref{sec: non_monotonic}). These results include a description of the upper and lower bounds on the resulting mean detection time, a sufficient condition for observing Zeno dynamics at a given measurement frequency, and bounds on the optimal measurement frequency and rate. The results are illustrated numerically with an example of a system of optically trapped ultra-cold atoms introduced in Section~\ref{sec: num_details}.

\begin{figure}[t]
\includegraphics[width=0.8\linewidth]{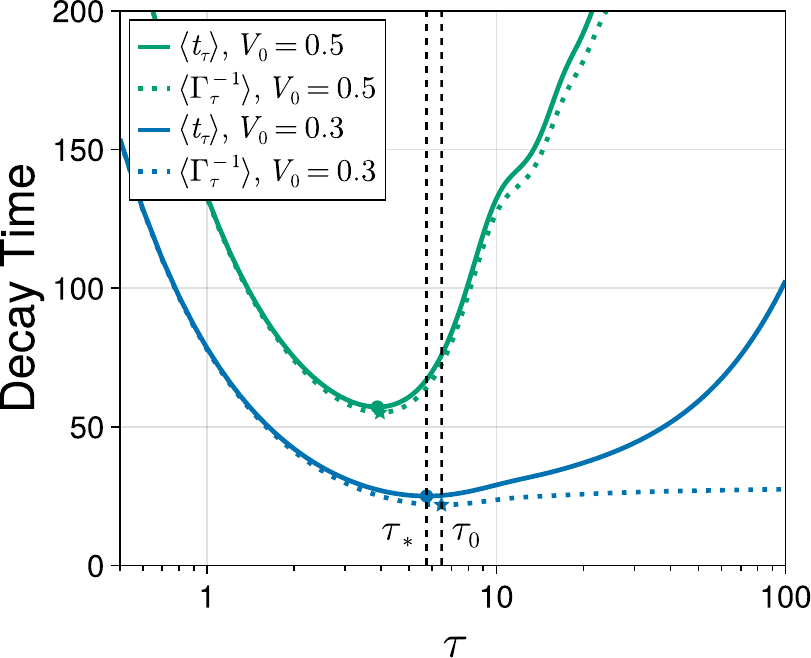}
\captionsetup{format=plain, justification=centerlast}
\caption{Characteristic escape time of quantum system from its initial unstable state in the presence of repeated measurements can be quantified either by the mean time required to detect decay $\langle t_\tau\rangle$ or by the inverse effective decay rate $\Gamma_\tau^{-1}$. The plot compare behavior of these quantities as the functions of measurement period $\tau$ in the case of optically-trapped atoms for two values of $V_0$ and fixed $a=0.1$. Generally, when $P(T)$ is arbitrary, one obtains approximate identity $\Gamma_\tau^{-1}\approx \langle t_\tau\rangle$ for sufficiently small values of $\tau$ and inequality $\Gamma_{\tau}^{-1}  \le  \langle t_\tau\rangle$  for all $\tau$ (see Eq.~(\ref{ordering_gamma_and_t})). If  the Zeno-inverse Zeno transition points are not multiple one additionally has $\tau_0>\tau_\ast$ (see Eq.~(\ref{relation1})). Clearly, $\Gamma_{\tau_0}^{-1} \le   \langle t_{\tau_\ast}\rangle$ (see Eq.~(\ref{ordering_gamma_and_t_ast})).}
\label{fig:1}
\end{figure}

\begin{figure*}[t]
\includegraphics[width=0.8\linewidth]{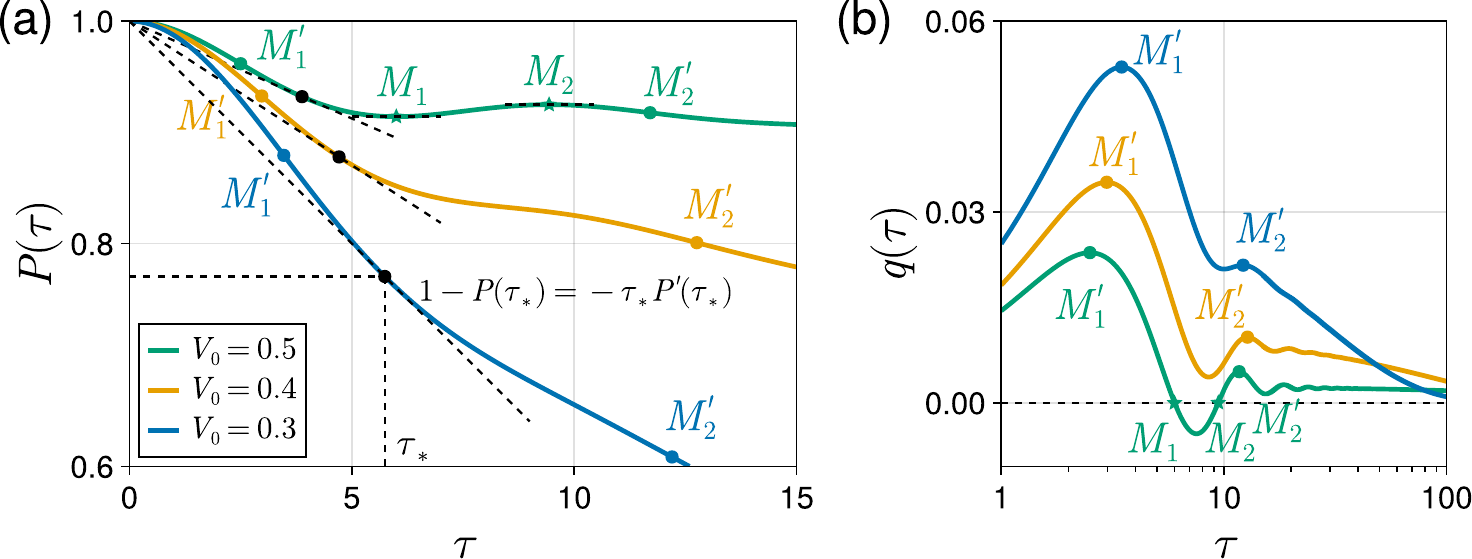}
\captionsetup{format=plain, justification=centerlast}
\caption{Survival probability $P(\tau)$ and its time derivative $q(\tau)=-dP(\tau)/d\tau$ for the system of optically-trapped cold atoms at different values of potential amplitude $V_0$ and fixed acceleration $a=0.1$. For sufficiently small values of $V_0$ the survival function $P(T)$ is monotonic so that $q(T)$ can be considered as probability density function of some non-negative random variable $T$. The plot also represents the positions of extrema  $M_1,M_2,\dots$ of $P(\tau)$, positions of maximums $M_1',M_2',\cdots$ of $q(\tau)$, and  points $\tau_\ast$ of Zeno - inverse Zeno transitions  captured by the properties of the mean decay detection time $\langle t_\tau\rangle$ under the stroboscopic measurement protocol with period $\tau$. Since the extrema condition for mean detection time is $- \tau_\ast\partial_\tau P(\tau_\ast)=1-P(\tau_\ast)$, as follows from Eq. (\ref{mean_detection_time_str}),  graphically the Zeno - inverse Zeno transition corresponds to the point in which the line passing through $(0,1)$ touches the graph of $P(\tau)$.
}
\label{fig:2}
\end{figure*}

Note that the majority of previous theoretical studies of Zeno dynamics have considered the effective decay rate of the survival probability as the key quantity of interest, rather than the mean detection time. For the stroboscopic measurement protocol, the inverse effective decay rate is given by
\begin{equation}
\label{inv_decay_rate_str}
   \Gamma_\tau^{-1}=-\frac{\tau}{\ln P(\tau)},
\end{equation}
see, e.g., Refs.~\cite{koshino2005quantum,chaudhry2016general,zhang2018criterion}. Zeno and inverse Zeno dynamics correspond to the conditions $\partial_\tau \Gamma_\tau^{-1}<0$ and $\partial_\tau \Gamma_\tau^{-1}>0$, respectively~\cite{kofman2000acceleration,chaudhry2016general,chaudhry2014zeno,segal2007zeno,thilagam2010zeno}. An example of a system exhibiting several Zeno-inverse Zeno transition points, i.e., points at which $\partial_\tau \Gamma_\tau^{-1}=0$, is described in Ref.~\cite{chaudhry2016general}. These two approaches to quantifying the state lifetime -- via the mean detection time or the inverse effective decay rate of the survival probability -- yield numerically similar results when the measurement period $\tau$ is sufficiently small so that $1-P(\tau)\ll 1$. In this regime, $\ln P(\tau)\approx P(\tau)-1$ and, consequently, $\Gamma_\tau^{-1}\approx \langle t_\tau\rangle$, see Fig.~\ref{fig:1}. The stroboscopic protocol outperforms all stochastic measurement protocols with independent and identically distributed intervals between measurements in terms of inverse effective decay rate optimization~\cite{belan2020optimality}. We denote by $\tau_0$ the optimal measurement period at which the inverse effective decay rate defined by Eq.~(\ref{inv_decay_rate_str}) attains its global minimum. It should be noted, however, that, unlike $\langle t_\tau\rangle$, the quantity $\Gamma_\tau^{-1}$ need not exhibit a non-monotonic dependence on $\tau$ and can be monotonic, as illustrated by the examples in Ref.~\cite{facchi2001quantum}. Here, we focus primarily on the behavior of $\langle t_\tau\rangle$, which always exhibits a Zeno-inverse Zeno transition point.

\section{Illustrative examples}
\label{sec: num_details}

To illustrate theoretical results derived further in this work, we consider ultra-cold atoms trapped in an accelerating periodic optical potential $V_0\cos(2k_Lx-k_Lat^2)$, where $V_0$ is the potential amplitude, $k_L$ is the laser wavenumber, and $a$ is the acceleration~\cite{fischer2001observation, wilkinson1997experimental}. In the accelerating reference frame $x' = x - at^2/2$, the potential can be written as $V_0\cos(2k_L x') + M a x'$, where $M$ is the atomic mass and the second term represents the inertial force acting on the atoms. 

\begin{table*}[t]
\caption{Zeno dynamics characteristics for the system of optically-trapped  atoms at different potential amplitudes $V_0$ and accelerations $a$. 
The survival probability $P(T)$ of the trapped state is given by Eq. (\ref{eq:surv_prob}).}
\begin{ruledtabular}
\begin{tabular}{c c c c c c c c c c c c c}
$V_0$ & $a$ & monotonic $P(T)$ & $\langle T^3 \rangle^{1/3}$ & $\tau_\ast$ & $t_{\tau_\ast}$ & $\tau_0$ & $\Gamma_{\tau_0}^{-1}$ & $1/r_\ast$ & $t_{r_\ast}$ & $h$ & $m_{\min}$ & $M_1'$ \\
\hline
0.6 & 0.1 & no  & -- & 3.27 & 80.3 & 3.31 & 78.7 & 1.55 & 110  & -- & 1469 & 2.13 \\
0.55 & 0.1 & no  & -- & 3.55 & 68.2 & 3.61 & 66.4 & 1.73 & 92.6 & -- & 577  & 2.30 \\
0.5  & 0.1 & no  & -- & 3.88 & 57.1 & 3.96 & 55.1 & 1.97 & 76.6 & -- & 244  & 2.50 \\
0.45 & 0.1 & no  & -- & 4.27 & 47.3 & 4.39 & 45.1 & 2.29 & 62.2 & -- & 111  & 2.72 \\
0.4  & 0.1 & yes & 158 & 4.71 & 38.6 & 4.92 & 36.2 & 2.75 & 49.6 & 11.1 & 54.7 & 2.97 \\
0.35 & 0.1 & yes & 85.2 & 5.22 & 31.2 & 5.59 & 28.4 & 3.34 & 38.9 & 8.7  & 29.5 & 3.23 \\
0.3  & 0.1 & yes & 49.7 & 5.74 & 25.0 & 6.46 & 21.8 & 3.91 & 30.5 & 7.0  & 17.6 & 3.47 \\
0.25 & 0.1 & yes & 31.4 & 6.16 & 20.1 & 7.72 & 16.5 & 4.15 & 24.2 & 5.8  & 11.7 & 3.63 \\
0.4 & 0.2 & yes & 20.9 & 3.96 & 12.3 & 4.83 & 9.93 & 2.56 & 15.0 & 3.6  & 7.3 & 2.37 \\	
0.4	& 0.15 & yes & 43.4 & 4.38 & 19.2 & 4.87 & 16.79 & 2.84 & 23.7 & 5.5 & 14.8 & 2.68 \\
0.4 & 0.05 & no & -- & 4.91 & 143 & 4.96 & 140 & 2.37 & 195 & -- & 1564 & 3.20 \\
\end{tabular}
\end{ruledtabular}
\label{table 1}
\end{table*}

Without acceleration, the system has Bloch energy bands separated by gaps; initially the lowest band is uniformly occupied while higher bands are empty. When acceleration is applied, the quasimomentum evolves through Bloch oscillations, and atoms may escape the lattice via interband Landau-Zener tunneling. The survival probability $P(T)$ in the trapped state after time $T$ can be calculated analytically~\cite{wilkinson1997experimental, niu1998landau}
\begin{equation}
\label{eq:surv_prob}
\ln P(T) = - \int_{0}^{T} d \tau (T-\tau) W(\tau),
\end{equation}
where
\begin{equation}
\begin{aligned}
   W(\tau) = \displaystyle \dfrac{a^2}{2 V_0} \int_{-\infty}^{\infty} ds \dfrac{1}{1+(s-a\tau/V_0)^2} \dfrac{1}{1+s^2}&\\
   \displaystyle \times \cos \left( \dfrac{V_0^2}{a} \int_{s}^{s-a\tau/V_0} dz \sqrt{1+z^2} \right),&
\end{aligned}
\end{equation}
and hereinafter we measure the time $T$ in units of $M/(4 \hbar k_L^2 )$, the potential amplitude $V_0$ in $4 \hbar^2 k_L^2/M$, and acceleration $a$ in $8 \hbar^2 k_L^3 /M^2$. The expression is valid for $\pi V_0^4/4 \lesssim a \lesssim \pi V_0^2/2$ and $0 < V_0 < 1$. 
The experimental technique intended to repeatedly measure the amount of trapped atoms was developed in Ref.~\cite{fischer2001observation}, and it was also demonstrated that the analytical expression (\ref{eq:surv_prob}), combined with the assumption of instantaneous and ideal measurements, yields results in good agreement with the experimental observations.

The considered system exhibits rich dynamics depending on the dimensionless parameters $V_0$ and $a$. First, we fix $a = 0.1$ and vary the potential amplitude within the range $0.25 \leq V_0 \leq 0.6$. Next, we set $V_0 = 0.4$ and vary the potential acceleration in the interval $0.05 \leq a \leq 0.2$. Qualitatively, the survival probability $P(T)$ behaves as a monotonically decreasing function with an exponential asymptotic decay at large $T$ for smaller values of $V_0 \leq 0.4$ and larger values of $a \geq 0.1$, and it exhibits an oscillatory behavior for the opposite parameter regimes, see also Fig.~\ref{fig:2}. 

The key quantities used in our analysis were computed numerically based on Eq.~(\ref{eq:surv_prob}) and are summarized in Table~\ref{table 1}. Integrals involving $q(T)=-dP(T)/dT$ are evaluated after applying integration by parts, which expresses them directly in terms of $P(T)$ and improves numerical stability by avoiding differentiation noise. The function $P(T)$ is precomputed on a dense grid of $T$ values, with intermediate points obtained via linear interpolation. For large times, $T \gg 1$, the interpolation is supplemented by the exponential asymptotic behavior of $P(T)$.

\section{Lower bound for the optimal mean detection time}
\label{sec: bottom bound}

As discussed in the Introduction, a carefully chosen interval between repeated measurements $\tau_\ast$ allows one to minimize the mean decay detection time of a given quantum state. Assume that $P(T)$ is monotonic and that $P(T) \to 0$ as $T \to \infty$. Then the function $F(T) = 1 - P(T)$ can be thought of as a cumulative distribution function for some non-negative random variable $T \geq 0$, which has the probability density function $q(T)=-dP/dT$ (see Fig. \ref{fig:2}). Given $F(\tau) \equiv Pr[T\le \tau]$, we can rewrite Eq.~(\ref{mean_detection_time_str}) as 
$\langle t_{\tau}\rangle=\tau/Pr[T \leq \tau]$.
Next, we can use Markov's inequality to obtain an upper bound on $Pr[T \leq \tau]=Pr[T^{-1} \geq \tau^{-1}] \leq \tau \langle T^{-1}\rangle$, and therefore
\begin{equation}
\langle t_{\tau}\rangle\ge h,
\end{equation}
where we have introduced the harmonic mean $h= \langle T^{-1} \rangle^{-1} =  \left(\int_0^\infty dT q(T)/T  \right)^{-1}$ of $T$. In particular, this means that the minimum mean time to decay also satisfies 
\begin{equation}
\label{lower_bound_res_h}
\langle t_{\tau_\ast}\rangle\ge h.
\end{equation}

Because the periodic measurement strategy with period $\tau_\ast$ is globally optimal (i.e., minimizes the mean decay time among all possible measurement strategies), we also have
\begin{equation}
\label{bottom_bound}
\langle t_{\cal M}\rangle \ge \langle t_{\tau_\ast}\rangle \ge h
\end{equation}
where ${\cal M}$ is an arbitrary measurement schedule. 
Thus, the mean decay detection time under an arbitrary measurement procedure is bounded from below by the harmonic mean associated with the survival probability function $P(T)$ assuming that the latter is monotonic.

\section{Upper bound for the optimal mean detection time}
\label{sec: upper bound}

\begin{figure*}[t]
\includegraphics[width=\linewidth]{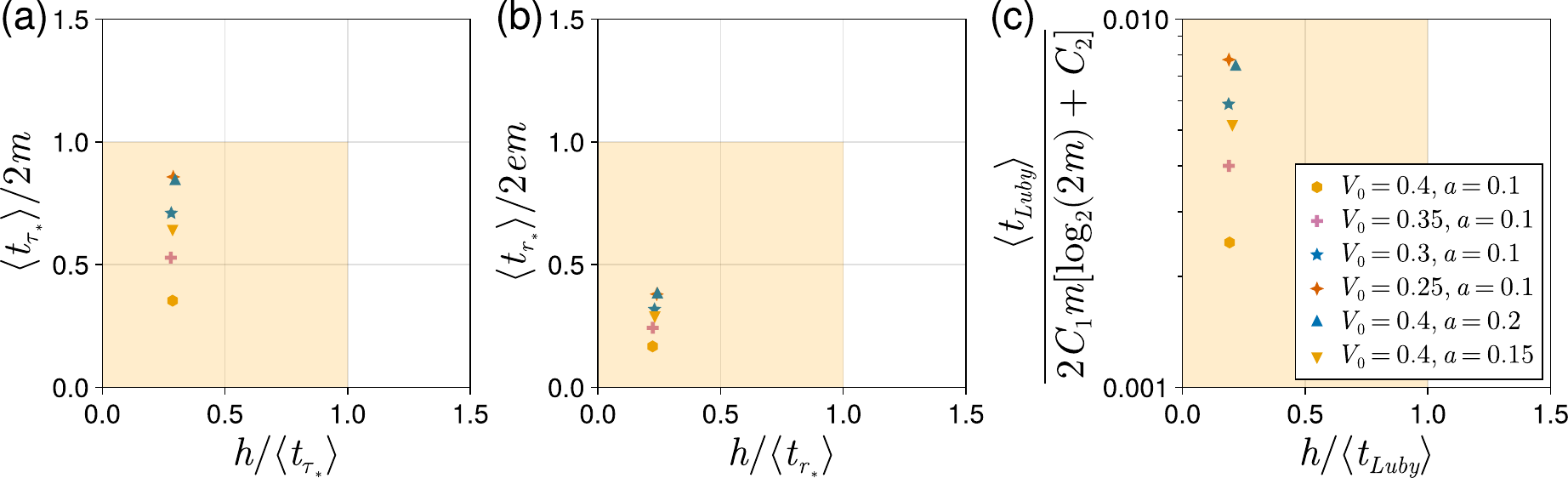}
\captionsetup{format=plain, justification=centerlast}
\caption{(a) Diagram in the plane of dimensionless parameters $x= h/\langle t_\tau\rangle$, $y=\langle t_{\tau_\ast}\rangle/(2m)$ illustrating relation between the optimized mean decay detection time $\langle t_{\tau_\ast}\rangle$ and the characteristic time scale of the  survival probability function $P(T)$: harmonic mean value $h$ and the  median $m$ of $q(T)= -dP(T)/dT \geq 0$ (see Eq. (\ref{lower_bound_res_h}) and (\ref{upper_bound_str})). (b) Diagram for the  optimized mean decay detection time $\langle t_{r_\ast}\rangle$  under optimally tuned Poisson measurement protocol. Here the vertical axis represents ratio $y=\langle t_{\tau_\ast}\rangle/(2em)$, see Eq. (\ref{mean_stoch_upper}). (c) Diagram illustrating the bounds (\ref{bound_universal_measurements}) and $\langle t_{Luby}\rangle\ge h$ (see Eq. (\ref{bottom_bound})) on the mean decay detection time $\langle t_{Luby}\rangle$ in the presence of Luby measurement protocol with $t_0=1$.}
\label{diagrams_for_str_and_stoch}
\end{figure*}

Since $\langle t_{\tau_\ast}\rangle\le \langle t_{\tau}\rangle$ for any $\tau$ by the definition of optimal period $\tau_\ast$, choosing $\tau=m$, where $m$ is defined by the condition $P(m)=1/2$, we find from Eq. (\ref{mean_detection_time_str}) that the optimized mean decay detection time does not exceed twice the half-life time of the quantum state: 
\begin{equation}
\label{upper_bound_str}
     \langle t_{\tau_\ast}\rangle\le 2m.
\end{equation}
Figure \ref{diagrams_for_str_and_stoch}a illustrates the bounds (\ref{lower_bound_res_h}) and (\ref{upper_bound_str}) for a system of cold atoms in accelerating optical potential for a range of values of $V_0$ (see Section \ref{sec: num_details} for details). We see that all point are indeed located in the region determined by the conditions $h/\langle t_{\tau_\ast}\rangle\le 1$ and $\langle t_{\tau_\ast}\rangle/(2m)\le 1$.

Next, setting $\tau=m$ in Eq.~(\ref{inv_decay_rate_str}), we obtain
\begin{equation}
\label{bound_gamma_tau0_m}
 \Gamma_{\tau_0}^{-1} \le \frac{m}{\ln 2},
\end{equation}
where $\tau_0$ denotes the optimal measurement period for minimizing the inverse decay rate, i.e., $\Gamma_{\tau_0}^{-1} \le \Gamma_\tau^{-1}$ for any $\tau$. We also note that
\begin{equation}
\label{ordering_gamma_and_t}
\Gamma_\tau^{-1} \le \langle t_\tau\rangle
\end{equation}
for any $\tau$. This inequality follows from Eqs.~(\ref{mean_detection_time_str}) and (\ref{inv_decay_rate_str}), together with the inequality $-\ln P(T)\ge 1-P(T)$. Consequently,
\begin{equation}
\label{ordering_gamma_and_t_ast}
\Gamma_{\tau_0}^{-1} \le \Gamma_{\tau_\ast}^{-1} \le \langle t_{\tau_\ast}\rangle,
\end{equation}
where the first inequality follows from the optimality of $\tau_0$ and the second from Eq.~(\ref{ordering_gamma_and_t}) (see also Eq.~(\ref{relation1}) below).

We now derive an upper bound on the mean decay detection time for a quantum state with survival probability $P(T)$ under a stochastic measurement protocol in which measurements occur at random times according to a Poisson process with rate parameter $r$. Defining the Laplace transform of $P(T)$ as $\tilde P(r)=\int_0^\infty e^{-rT}P(T)dT$, integration by parts gives $\tilde P(r) = [1-\tilde q(r)]/r$, where $q(T)=-dP(T)/dT \geq 0$ if $P(T)$ is monotonic. Consequently, the mean detection time, given by Eq.~(\ref{poisson_mdt}), can be written as
\begin{equation}
\label{mean_detection_Poisson}
    \langle t_r\rangle=\frac{1}{r\tilde q(r)}.
\end{equation}
The optimal measurement rate $r_\ast$ is defined by the condition $\langle t_{r_\ast}\rangle \leq \langle t_r\rangle$ for any $r$. Choosing $r=m^{-1}$, where $m$ is defined by $P(m)=1/2$, and using Eq.~(\ref{mean_detection_Poisson}), we obtain
\begin{equation}
\begin{aligned}\label{mean_stoch_upper}
\langle t_{r_\ast}\rangle  \le   \langle t_{m^{-1}}\rangle=\frac{m}{\int_0^\infty dT q(T)e^{-T/m}} \le \\
\le \frac{m}{\int_0^m dT q(T)e^{-T/m}}\le \frac{m}{e^{-1}\int_0^m dT q(T)}=2em.
\end{aligned}
\end{equation}
The upper bound in Eq.~(\ref{mean_stoch_upper}), together with the lower bound $\langle t_{r_\ast}\rangle\geq h$, which follows from Eq.~(\ref{bottom_bound}), is illustrated in Fig.~\ref{diagrams_for_str_and_stoch}b.

The stroboscopic protocol with period $\tau_\ast$ always yields the shortest expected decay detection time among all measurement strategies, although the optimal period depends on the details of the system. It has previously been shown that the so-called Luby measurement protocol can achieve a mean decay detection time relatively close to the optimal value $\langle t_{\tau_\ast}\rangle$, even when the survival probability $P(T)$ is unknown. The Luby protocol is characterized by the following sequence of inter-measurement intervals, $\{T_1, T_2, T_3,...\}$, defined by $T_i=2^{k-1}t_0$ if $i=2^k-1$, and $T_i=T_{i-2^{k-1}+1}$ if $2^{k-1}\le i< 2^k-1$. The time step $t_0$ should be small compared with the characteristic decay time scale. The beginning of this sequence is $t_0$, $t_0$, $2t_0$, $t_0$, $t_0$, $2t_0$, $4t_0$, $t_0$, $t_0$, $2t_0$, $t_0$, $t_0$, $2t_0$, $4t_0$, $8t_0$, $t_0$, $\dots$. Rather than relying on a single, potentially suboptimal measurement period, the Luby strategy explores a range of inter-measurement intervals, allocating approximately equal fractions of the total time to attempts using each interval. The mean decay detection time under this protocol satisfies the bound $\langle t_{Luby}\rangle\le C_1\langle t_{\tau_\ast}\rangle (\log_2 \frac{\langle t_{\tau_\ast}\rangle}{t_0}+C_2)$, where $C_1=\frac{8}{(1-e^{-1})^2} \approx 20.0$ and $C_2=(1-e^{-1})^2\sum_{k=1}^\infty k e^{-k+1} \log_2k + 3 \approx 3.87$ \cite{belan2020optimality}. This bound can be derived from the general expression $\langle t\rangle=T_1+\sum_{n=2}^\infty T_n\prod_{k=1}^{n-1}P(T_k)$ for the expected decay detection time under an arbitrary deterministic measurement protocol ${\cal M}$, specified by the sequence of inter-measurement intervals $T_1, T_2, T_3,\dots$. The derivation relies on probabilistic arguments that exploit the assumed monotonicity of the survival probability $P(T)$ and the specific structure of the Luby sequence. Combining this result with Eq.~(\ref{upper_bound_str}), we obtain
\begin{equation}
\label{bound_universal_measurements}
    \langle t_{Luby}\rangle\le 2C_1m( \log_2 \frac{2m}{t_0}+C_2).
\end{equation}
This result bounds the performance of the Luby strategy in terms of the characteristic time scale $m$ of the survival probability $P(T)$, without explicitly involving $\langle t_{\tau_\ast}\rangle$, whose calculation requires detailed knowledge of the functional form of $P(T)$. The upper bound in Eq.~(\ref{bound_universal_measurements}), together with the lower bound $\langle t_{\mathrm{Luby}}\rangle\geq h$, which follows from Eq.~(\ref{bottom_bound}), is illustrated in Fig.~\ref{diagrams_for_str_and_stoch}c.

\section{Sufficient conditions for observing direct/inverse Zeno dynamics}
\label{sec: suff_cond}

Once the survival probability $P(T)$ is known, Eq.~(\ref{mean_detection_time_str}) can be used to determine whether measurements performed at a given frequency induce the quantum Zeno or inverse quantum Zeno effect. Furthermore, simple criteria for identifying different decay regimes can be formulated in terms of integral or local properties of $P(T)$. In particular, as illustrated in Fig.~\ref{fig:2}, stroboscopic measurements with period
\begin{equation}
\tau\le M_1',
\end{equation}
where $M_1'>0$ denotes the location of the leftmost maximum of $q(T)=-dP(T)/dT$, always lead to the direct quantum Zeno regime, characterized by $\partial_\tau \langle t_\tau\rangle <0$.

For Poissonian measurements, we use the inequalities $\tilde q(r)\ge 1-r\langle T\rangle +\frac{r^2}{2}\langle T^2\rangle-\frac{r^3}{6}\langle T^3\rangle$ (see Ref.~\cite{zubkov1999two} and Eq.~(20) of Ref.~\cite{nikitin2024constructing}) and $\partial_r \tilde q(r)\ge -\langle T\rangle +r \langle T^2\rangle-\frac{r^2}{2}\langle T^3\rangle$ (see Eqs.~(20) and (29) of Ref.~\cite{nikitin2024constructing}). Here, $\langle T^n\rangle=\int_0^\infty dT T^n q(T)$ where, as assumed above, $q(T)\geq 0$ for all $T\geq 0$ and is normalized according to $\int_0^\infty q(T) dT=1$. Combining these inequalities, we obtain the following estimate:
\begin{align*} 
\label{estimate0}
\partial_r \langle t_r\rangle^{-1}=\tilde q(r)+r\partial_r\tilde q(r)\ge 1-2r\langle T\rangle + \\
+\frac{3}{2}r^2\langle T^2\rangle-\frac{2}{3}r^3\langle T^3\rangle>  1-2r\langle T\rangle  
-\frac{2}{3}r^3\langle T^3\rangle>\\
>1-2r\langle T^3\rangle^{1/3}  
-\frac{2}{3}r^3\langle T^3\rangle.
\end{align*}
Consequently, the system is in the inverse quantum Zeno regime whenever the measurement rate is sufficiently small, namely,
\begin{equation}
\label{suff_condition_recerse_zeno}
    r<(2^{1/3}-2^{-1/3})\langle T^3\rangle^{-1/3},
\end{equation}
which ensures that $\partial_r \langle t_r\rangle^{-1}> 1-2r\langle T^3\rangle^{1/3}  
-\frac{2}{3}r^3\langle T^3\rangle>0$.

A necessary and sufficient condition for the inverse Zeno effect in the Poissonian measurement protocol, expressed in terms of the first two moments of the detection time, $\langle t_r\rangle$ and $\langle t_r^2\rangle$, was previously established in Ref.~\cite{belan2020optimality}. Specifically, the inequality $\langle t_r^2\rangle>2\langle t_r\rangle^2-2r^{-1} \langle t_r\rangle$ is equivalent to the condition $\partial_r\langle t_r\rangle<0$, which characterizes the inverse Zeno effect under Poissonian measurements. In contrast, criterion (\ref{suff_condition_recerse_zeno}) involves only a single integral characteristic of the survival function $P(T)$, rather than two temporal moments of the detection-time distribution under the measurement protocol. However, it is important to emphasize that inequality (\ref{suff_condition_recerse_zeno}) provides a sufficient, but not necessary, condition for the occurrence of inverse Zeno dynamics. Consequently, the range of measurement rates for which the inverse Zeno effect occurs may extend beyond the range specified by this inequality. For model-specific criteria that relate the occurrence of Zeno or inverse Zeno dynamics to the details of the system Hamiltonian, see Refs.~\cite{facchi2001quantum,koshino2005quantum,zhang2018criterion}.

\section{Lower bounds for  optimal measurement periods and rates}
\label{sec: estimates for optimal}

Using the extremum condition $ - \tau_\ast\partial_\tau P(\tau_\ast)=1-P(\tau_\ast)$ for $\langle t_\tau\rangle$, which follows from Eq.~(\ref{mean_detection_time_str}), and referring to Fig.~\ref{fig:2}, we conclude that the optimal measurement period $\tau_\ast$, which minimizes the mean decay detection time, satisfies
\begin{equation}
\label{bottom_bound1}
\tau_\ast > M_1',
\end{equation}
where, as before, $M_1'$ denotes the location of the leftmost maximum of $q(T)=-dP(T)/dT$.

\begin{figure}[t]
\includegraphics[width=0.7\linewidth]{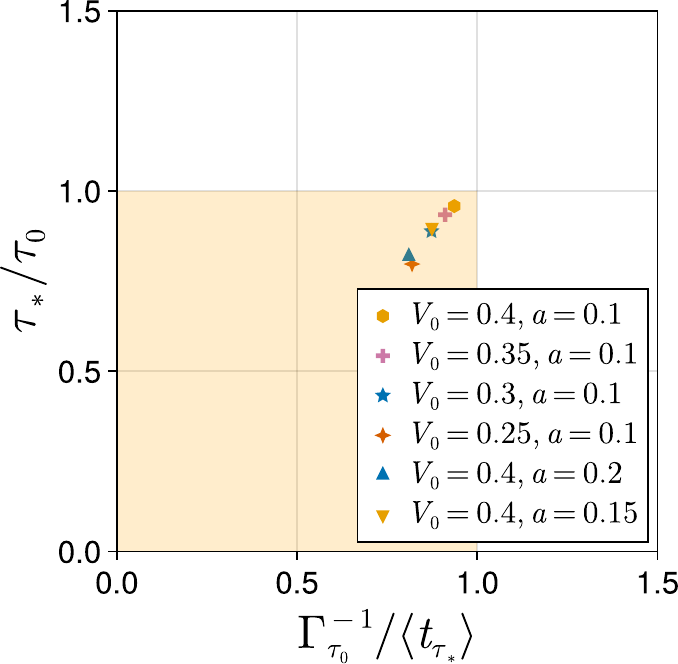}
\captionsetup{format=plain, justification=centerlast}
\caption{Diagram in the plane of dimensionless parameters $x= \Gamma_{\tau_0}^{-1}/\langle t_{\tau_\ast}\rangle$, $y=\tau_\ast/\tau_0$ for the system of optically-trapped cold atoms at different values of potential amplitude $V_0$ and  acceleration $a$. As discussed in the text, the optimal periods  of the Zeno-inverse Zeno transition and the resulting state lifetimes quantified from two perspectives -- via the mean decay detection time and the inverse effective decay rate -- obeys the inequalities $\Gamma_{\tau_0}^{-1} \le \langle t_{\tau_\ast}\rangle$ and $\tau_0>\tau_\ast$ (when the transition points are not multiple), see Eqs.~(\ref{ordering_gamma_and_t_ast}) and (\ref{relation1}).}
\label{fig: rate and detection time}
\end{figure}

Assume that there is a single transition between the Zeno and inverse Zeno regimes, both in terms of the mean detection time $\langle t_\tau\rangle$ and the effective decay rate $\Gamma_\tau$. When the characteristic timescale of the dynamics is quantified by the effective decay rate, the direct Zeno regime is characterized by $\partial_\tau\Gamma_\tau=\frac{-\tau\partial_\tau P(\tau)+P(\tau)\ln P(\tau)}{\tau^2 P(\tau)}>0$. Consequently, at the optimal measurement period $\tau_\ast$ that minimizes the mean detection time, we have
\begin{equation} 
\partial_\tau\Gamma_{\tau}|_{\tau_\ast}=\frac{1- P(\tau_\ast)+P(\tau_\ast)\ln P(\tau_\ast)}{\tau^2_\ast P(\tau_\ast)}
>0,
\end{equation}
where we used the extremum condition for $\langle t_\tau\rangle$ and the fact that $0<P(\tau_\ast)\le 1$. This implies that the optimal measurement period $\tau_0$ associated with the effective decay rate exceeds $\tau_\ast$:
\begin{equation}
\label{relation1}
\tau_0>\tau_\ast,
\end{equation}
as illustrated in Fig.~\ref{fig: rate and detection time}. Combining this result with Eq.~(\ref{bottom_bound1}), we obtain the lower bound
\begin{equation}
\label{bound_tau0}
\tau_0> \tau_\ast > M_1'.
\end{equation}

\begin{figure}[t]
\includegraphics[width=0.7\linewidth]{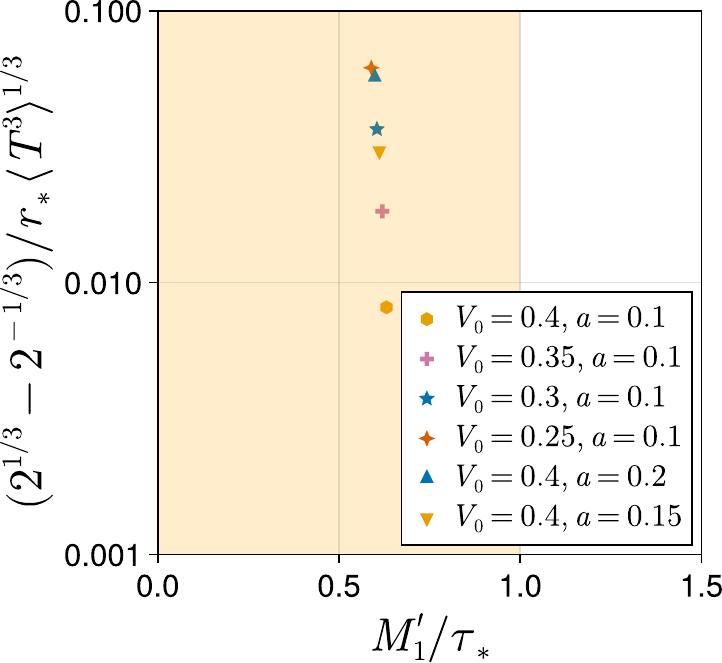}
  \captionsetup{format=plain, justification=centerlast}
  \caption{Diagram illustrating the bounds (\ref{bottom_bound1}) and (\ref{bound_r_ast}) on the optimal stroboscopic period $\tau_{\ast}$ and optimal Poisson measurement rate $r_{\ast}$ for the system of optically-trapped cold atoms at different values of potential amplitude $V_0$ and acceleration $a$.}
  \label{fig_5}
\end{figure}

Next, once the third order moment $\langle T^3\rangle=\int_0^\infty dT  q(T)T^3$ is known, a lower bound on the optimal stochastic measurement rate $r_\ast$ can be established. Specifically, the sufficient condition for the occurrence of inverse Zeno dynamics, given by inequality (\ref{suff_condition_recerse_zeno}), yields
\begin{equation}
\label{bound_r_ast}
r_\ast\ge (2^{1/3}-2^{-1/3})\langle T^3\rangle^{-1/3}.
\end{equation}
Numerical examples illustrating the validity of the bounds in Eqs.~(\ref{bottom_bound1}), (\ref{bound_tau0}), and (\ref{bound_r_ast}) are provided in Table~\ref{table 1} and Fig.~\ref{fig_5}.

An additional lower bound on the optimal stochastic measurement rate $r_\ast$ can be obtained using the relation $\langle t_{r_\ast}^2\rangle=2\langle t_{r_\ast}\rangle^2-2r_\ast^{-1}\langle t_{r_\ast}\rangle$ \cite{belan2020optimality}, together with the general inequality $\langle t_r^2\rangle\ge\langle t_r\rangle^2$ and Eq.~(\ref{mean_stoch_upper}). Specifically,
\begin{eqnarray}
\langle t_{r_\ast}\rangle^2\le     \langle t_{r_\ast}^2\rangle=2\langle t_{r_\ast}\rangle^2-2r^{-1}_\ast\langle t_{r_\ast}\rangle,
\end{eqnarray}
which implies
\begin{equation}
\label{bound_r_ast_1}
    r_\ast\ge 2\langle t_{r_\ast}\rangle^{-1} \ge e^{-1}m^{-1}.
\end{equation}

Notably, the lower bound in Eq.~(\ref{bound_r_ast_1}), which relates the optimal measurement rate $r_\ast$ to the median value $m$, can be improved. To this end, consider measurement rates satisfying $rm<1$. We have
\begin{eqnarray*}
&&\frac{d}{dr}\frac{1}{\langle t_r\rangle}=\frac{d}{dr}(r\tilde{q}(r))=\\
&&=\frac{d}{dr}(r\int_0^\infty dT q(T)e^{-rT})=\\
&&=\int_0^\infty dT q(T)(1-rT)e^{-rT}=\\
&&=\int_0^{m} dT q(T)(1-rT)e^{-rT}+\\
&&+\int_m^{\infty} dT q(T)(1-rT)e^{-rT}\ge \\
&&\ge (1-rm)e^{-rm}\int_0^{m} dT q(T)+\\
&&+min_{x} [(1-x)e^{-x}]\int_m^{\infty} dT q(T)=\\
&&=\frac{1}{2}((1-rm)e^{-rm}-e^{-2}).
\end{eqnarray*}
Here, we have used the definition of the median, which implies \(\int_0^m q(T)\,dT=\int_m^\infty q(T)\,dT=1/2\), together with \(\min_{x\ge0}(1-x)e^{-x}=-e^{-2}\). It follows that the inverse Zeno effect occurs whenever
\begin{equation}
\frac{1}{2}((1-rm)e^{-rm}-e^{-2})>0,
\end{equation}
or, equivalently, $r<x_0m^{-1}$, where $x_0=1-W(e^{-1})\approx0.7215$, and $W(z)$ denotes the Lambert $W$-function~\cite{barry1995real}. The constant $x_0$ is the solution of
\[
(1-x_0)e^{-x_0}-e^{-2}=0.
\]
Consequently, the optimal measurement rate satisfies the improved lower bound
\begin{equation}
\label{improved_estimate}
r_\ast>0.721\,m^{-1}.
\end{equation}
The numerical constant in this bound is larger than the corresponding constant in Eq.~(\ref{bound_r_ast_1}).

\section{Non-monotonic survival probability}
\label{sec: non_monotonic}

So far, we have assumed that the survival probability $P(T)$ decreases monotonically and approaches zero as $T\to\infty$. We now relax this assumption and consider arbitrary survival probabilities, including non-monotonic profiles. Using the relation
\begin{equation}
\label{simple_res}
\langle t_{\tau_\ast}\rangle>  \tau_\ast,
\end{equation}
which follows directly from Eq.~(\ref{mean_detection_time_str}), together with Eq.~(\ref{bottom_bound1}), we obtain
\begin{equation}
\label{nonmonotonic1}
\langle t_{\tau_\ast}\rangle>M_1'.
\end{equation}
Here, $M_1'$ denotes the location of the leftmost local maximum of $q(T)=-dP(T)/dT$. Both inequalities remain valid for non-monotonic survival probabilities. Note that Eq.~(\ref{simple_res}) is analogous to the corresponding bound for optimal Poissonian measurements, $\langle t_{r_\ast}\rangle\ge 2r_\ast^{-1}$, established in Ref.~\cite{belan2020optimality}. Therefore, the mean decay detection time is strictly greater than the location of the leftmost local maximum of $-dP(T)/dT$. This bound is more general than Eq.~(\ref{lower_bound_res_h}), as its derivation does not require the survival probability $P(T)$ to be monotonic.

\begin{figure}[t]
\includegraphics[width=0.7\linewidth]{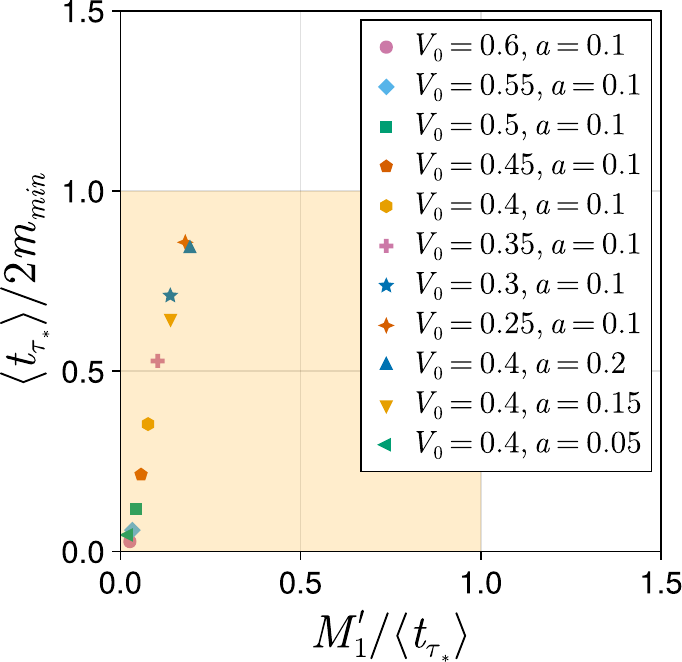}
\captionsetup{format=plain, justification=centerlast}
\caption{Diagram in the plane of dimensionless parameters $x=M_1'/\langle t_\tau\rangle$, $y=\langle t_{\tau_\ast}\rangle/(2m_{min})$ illustrating relation between the optimized mean decay detection time $\langle t_{\tau_\ast}\rangle$ and the characteristic time scales of the survival probability function $P(T)$: the leftmost median $m_{min}$ and position of the  leftmost maximum $M_1'$ of $-dP(T)/dT$ (see Eqs. (\ref{nonmonotonic1}) and (\ref{nonmonotonic2})).}
\label{fig: nonmonotonic}
\end{figure}

Regarding the upper bound (\ref{upper_bound_str}) discussed in Sec.~\ref{sec: upper bound}, for a non-monotonic survival probability $P(T)$, the median $m$ need not be unique. In this case, Eq.~(\ref{upper_bound_str}) should be replaced by
\begin{equation}
\label{nonmonotonic2}
\langle t_{\tau_\ast}\rangle\le 2m_{\min},
\end{equation}
where $m_{\min}$ is the smallest value of $T$ satisfying $P(m_{\min})=1/2$. The numerical results presented in Fig.~\ref{fig: nonmonotonic} confirm that both bounds (\ref{nonmonotonic1}) and (\ref{nonmonotonic2}) hold for a system of atoms in an accelerating periodic optical potential under optimal stroboscopic measurements.

Assume that there is a single transition point $\tau_\ast$ between the Zeno and inverse Zeno regimes (if multiple transition points exist, the inequality remains valid only for the leftmost transition point). The optimal measurement period $\tau_\ast$ then satisfies the another upper bound
\begin{equation}
\label{eq_res}
\tau_\ast<M_1,
\end{equation}
as follows from the extremum condition $-\tau_\ast\partial_\tau P(\tau_\ast)=1-P(\tau_\ast)$ for $\langle t_\tau\rangle$ and Fig.~\ref{fig:2}. Here, $M_1>0$ denotes the location of the leftmost nonzero extremum of $P(T)$. Furthermore, we note that any measurement period $\tau=M_1,M_2,\ldots$, where $0<M_1<M_2<\cdots$ denotes the set of all extrema of $P(T)$, lies in the inverse Zeno regime, characterized by the condition $\partial_\tau\langle t_\tau\rangle>0$.

A finite duration of the measurement events may modify the results obtained in this work, although some of the corresponding generalizations are straightforward. For example, under a stroboscopic measurement protocol with period $\tau$ and measurement duration $\Delta$, the mean decay detection time is given by $\langle t_\tau(\Delta )\rangle=\frac{\tau+\Delta}{1-P(T)}$~\cite{belan2020optimality}. Consequently, its optimal value satisfies the upper bound $\langle t_{\tau_\ast}(\Delta)\rangle\le 2m_{\min}+2\Delta$. The generalization of the remaining results to measurements of finite duration requires a separate analysis.

\section{Conclusion}

Zeno dynamics can be characterized in terms of statistical properties of the state survival probability $P(T)$, without explicit reference to system-specific physical parameters. This framework provides bounds on the mean decay detection time (Secs.~\ref{sec: bottom bound} and \ref{sec: upper bound}), estimates of the optimal measurement period and rate (Sec.~\ref{sec: estimates for optimal}), and sufficient conditions for the occurrence of the Zeno and inverse Zeno effects (Sec.~\ref{sec: suff_cond}). Our analysis encompasses three measurement protocols -- stroboscopic, Poissonian, and Luby -- and is illustrated numerically using a system of cold atoms in an accelerating periodic optical potential. Among the main results, we highlight the universal lower bound $\langle t_{\cal M}\rangle \ge h$ [Eq. (\ref{bottom_bound})], which applies to any quantum system and measurement protocol ${\cal M}$, provided that the survival function $P(T)$ is monotonic. Section~\ref{sec: non_monotonic} extends the analysis to non-monotonic survival functions. Although most of our analysis focuses on the mean decay detection time, several results also apply to the optimization of the effective decay rate associated with the state survival probability; see Eqs. (\ref{bound_gamma_tau0_m}), (\ref{relation1}), and (\ref{bound_tau0}).

The Zeno and inverse Zeno effects can also be characterized by comparing the effective decay rates with and without measurements, $\Gamma_\tau$ and $\Gamma$, respectively~\cite{zhang2018criterion}. The two approaches -- one based on the sign of $\partial_\tau\Gamma_\tau$ (or, equivalently, $\partial_\tau\langle t_\tau\rangle$) and the other on the ratio $\Gamma_\tau/\Gamma$ -- are not equivalent and may classify the dynamics of the same system under the same measurement protocol differently. Regardless of the criterion adopted to distinguish between the Zeno and inverse Zeno regimes, the present analysis establishes bounds on the effective decay rate $\Gamma_{\tau_0}$ at its optimal measurement period $\tau_0$; see Eqs.~(\ref{bound_gamma_tau0_m}) and (\ref{bound_tau0}).

\acknowledgments

The authors are grateful to the anonymous Referee for suggesting the arguments that led to Eq.~(\ref{improved_estimate}). Work of V.P. and S.B. on derivation of analytical results was supported by the Basic research program of HSE University (project number HSE-BR-2025-57). Work of V.P. and S.B. on numerical examples was supported by the Ministry of Science and Higher Education of the Russian Federation within the state assignment of the Landau Institute for Theoretical Physics.

\bibliography{biblio}

\end{document}